\documentclass[conference]{IEEEtran}
\IEEEoverridecommandlockouts
\usepackage{cite}
\usepackage{amsmath,amssymb,amsfonts}
\usepackage{algorithmic}
\usepackage{graphicx}
\graphicspath{ {figures/} }
\usepackage{textcomp}

\usepackage{subfigure}
\usepackage[table]{xcolor}
\newcommand\ChangeRT[1]{\noalign{\hrule height #1}}

\usepackage[colorinlistoftodos]{todonotes}

\def\BibTeX{{\rm B\kern-.05em{\sc i\kern-.025em b}\kern-.08em
    T\kern-.1667em\lower.7ex\hbox{E}\kern-.125emX}}
\begin{document}

\title{A2D: Anywhere Anytime Drumming \\
\thanks{\textsuperscript{*}Equal contribution}
}
\author{
\IEEEauthorblockN{Harel Yadid\textsuperscript{*}}
\IEEEauthorblockA{
\textit{Technion -- Israel Institute of Technology} \\
\textit{Faculty of Electrical and Computer Engineering} \\
Haifa, Israel \\
hyadid120@gmail.com}
\and
\IEEEauthorblockN{Almog Algranti\textsuperscript{*}}
\IEEEauthorblockA{
\textit{Technion -- Israel Institute of Technology} \\
\textit{Faculty of Computer Science}\\
Haifa, Israel \\
almog.alg@gmail.com}
\and
\IEEEauthorblockN{Mark Levin}
\IEEEauthorblockA{\textit{Technion -- Israel Institute of Technology} \\
\textit{Faculty of Electrical and Computer Engineering} \\
Haifa, Israel \\
levmarki@technion.ac.il}
\and
\IEEEauthorblockN{Ayal Taitler}
\IEEEauthorblockA{
\textit{University of Toronto} \\
\textit{Department of Mechanical and and Industrial Engineering} \\
Toronto, Ontario, Canada \\
ataitler@gmail.com}
}

\maketitle

\begin{abstract}
The drum kit, which has only been around for around 100 years, is a popular instrument in many music genres such as pop, rock, and jazz. However, the road to owning a kit is expensive, both financially and space-wise. Also, drums are more difficult to move around compared to other instruments, as they do not fit into a single bag. We propose a no-drums approach that uses only two sticks and a smartphone or a webcam to provide an air-drumming experience. The detection algorithm combines deep learning tools with tracking methods for an enhanced user experience. Based on both quantitative and qualitative testing with humans-in-the-loop, we show that our system has zero misses for beginner level play and negligible misses for advanced level play. Additionally, our limited human trials suggest potential directions for future research.
\end{abstract}


\section{Introduction} \label{sec:intro}

Learning a musical instrument requires significant time and money in order to excel. In addition, even to determine whether this is the right instrument for them, beginners must purchase a beginner level instrument up front. These instruments are not cheap either. The same thing happens when someone wants to start drumming. Typically, the funds are spent on drums, a tutor, and a large enough soundproof environment. To address these issues, at least for the non-professional player, the future envisions an air drumming system that will allow users to play anywhere, at any time, and at a low cost.
 However, tracking small objects, such as the tip of a drumstick, can be a challenging task. Aerodrums \cite{Aerodrums}, the air drumming industry's pioneer, bases its technology on the recognition of two stick markers and two leg markers. However, their method is light-dependent, necessitates specialized equipment, a camera, and only allows for air drumming. Another type of solution, the AeroBand PocketDrum \cite{Aeroband}, offers detector-based air drums without the use of cameras, though they are also limited to playing in the air and require specialized technology. 

Current commercial real-time systems such as Aerodrums and PocketDrum require dedicated hardware in order to operate. Even though specialized hardware has the potential to perform adequately for this specific purpose, it still hinders the ideal freedom of air drumming somewhat. It costs money, and in reality, it does not meet the performance expectations of such a system. Academia has also studied the problem. Several previous works have attempted to address hardware-free drumming. In \cite{tolentino2019air} attempted to apply classical vision methods such as blob detection to locate the drumsticks. \cite{Phon2009} tried to find the centroid of the tip of the drumsticks and track them using first order approximation of the movement between two following frames. \cite{Fidan2010} employed color segmentation to find the drumsticks, and adaptively adjusted the region of interest of the drumsticks based on a tracking filter. While the solutions above require specialized equipment, we propose using a component that is already in everyone's pocket, a smartphone.

We propose a completely no-drum approach in this work, relaxing the previous works' demands for dedicated hardware or even specialized markers on the drumsticks. Furthermore, Our approach enables users to play on physical surfaces with the necessary surface feedback to simulate a realistic drumming experience, as well as in the air, akin to current solutions. Recent advancements in deep learning have boosted computer vision performance significantly \cite{krizhevsky2017imagenet}. Image segmentation and object detection is sub-field of computer vision, which has seen major advances as well \cite{redmon2015real, tan2020efficientdet, Yolov7}. We build upon these advancements and fine-tune a pre-trained YOLOv5 neural network on a custom-made image dataset of air and other surfaces drumming people. As the image detection module might not be completely accurate, we track the tip of the drumsticks with a second order kinematic tracking Kalman filter \cite{Kalman1960}. We treat the tracking problem as a disappearing objects problem \cite{Kalmanaero}, and estimate a belief on the future position and velocity of the tip of the drumsticks. Thus, our solution is able to detect any generic drumsticks without any markers, with the use of a simple webcam or a smartphone. We provide experiments showing the accuracy of the tracking and missed hits for several play levels. Our result suggests that for 80 hits per minute and below, the performance is accurate and precise.

\section{Problem Statement} \label{sec:statement}

The problem considered here is drumming with generic, unmarked drumsticks in the air or any other surface as long as the drumsticks are not concealed, where a sound should be played whenever the drumsticks \textit{hit} a drum. A hit and its volume type and magnitude is denoted by a sharp movement change made in a specified designated area in the frame.
In this work, we used a drum kit with five distinct sounds, however, there is no limitation on the number of drums and appropriate sound our system supports.
 As the system is intended for real-time use by humans, the solution is expected to obey three requirements.
\begin{enumerate}
    \item Minimum delay between the time the player \textit{hits the drum} and the time the appropriate note is sounded.
    \item Minimum misses of drum hits, and with the appropriate volume and drum sound.
    \item This solution should be able to be executed on a standard smartphone or equivalent hardware. 
\end{enumerate}
In this case, the low hardware requirements are in direct competition with the requirements for minimum time and miss. The goal is to find a compromise such that the hardware will be simple enough while the human experience will be minimally affected. In this work, we assume a standard entry-level camera, present in mobile devices can take high-definition videos and is capable of at least $60$ frames per second (FPS).
Note, that while a zero delay between hit and detection is theoretically impossible, the human ear does not "feel" delays of approximately $15[ms]$ or less \cite{haas1972influence}. A 60 FPS rate is ${\sim} 16[ms]$ between frames, which is an acceptable compromise.

\section{Preliminaries}
The method we used relies heavily on deep learning tools. Before any detection can be made, a dedicated neural network was trained on data gathered specially for this task.

\begin{figure}
\centering
\includegraphics[width=0.45\textwidth]{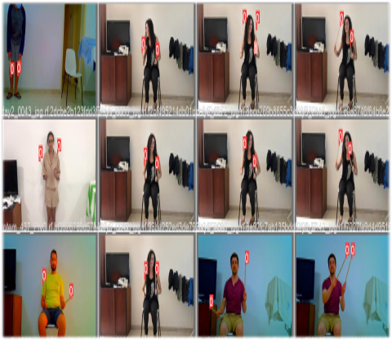}
\caption{A sample of images from the dataset used to train the detection network. Each image is annotated with squares denoting the locations of the tips of the drumsticks.}\label{fig:dataframes}
\end{figure}

As object detection networks are not trained for our specific type of images and are not required to detect a stick's tip specifically, we used a pre-trained network and fine-tuned it for images of people air playing drums with drumsticks. The images were extracted from recorded videos generated especially for that task. In order for the videos to be diverse enough to accommodate new users in unseen environments, we collected videos of a range of people and setups. We noted different genders, skin colors, rooms, room lighting, sitting setups and stances, and the camera calibration was different between different videos as well. In figure \ref{fig:dataframes} a sample of the frames extracted from a specific recorded setup can be seen. In addition, the collected data has passed through a series of processing procedures, such as changing images' brightness, changing their resolution, changing their orientation, etc. Finally, the data was filtered so duplicate or too similar images were removed, and augmented with freely available images of people playing from the internet. The total amount of images in the final dataset was 5000 images. This amount is insufficient in order to train from scratch a deep neural network, but it was enough for fine-tuning our network, as significant amount of the "knowledge" was transferred from the pre-trained network.
For each of the images, the tips of the drumsticks were marked manually.

\subsection{Data Collection and Network Training}
Recent years have brought significant advancements to the field of computer vision \cite{krizhevsky2017imagenet}, among them are image classification \cite{simonyan2014very, he2016deep} and object detection \cite{Yolov7, tan2020efficientdet}. The quality of these approaches is typically measured by their performance on benchmark datasets such as \cite{deng2009imagenet} for image classification and COCO \cite{lin2014microsoft} for object detection.

In this work, we addressed the task of detecting the drumsticks' tips by employing a pre-trained network which we fine-tuned on our dedicated dataset of people playing drums in the air or on a surface. As our dataset is comprised of natural images of people and everyday objects, taking such an approach with a pre-trained network reduced the training time significantly compared to training from scratch. The requirements in this work for the detection algorithm were speed, accuracy, and low memory and computational footprint, as the system is intended to run on a low-power device and not high-performance hardware such as GPUs. The chosen architecture was the small version of YOLOv5 \cite{yolo5}, for its relatively small footprint, high speed, and sufficient detection performance, also supported by the results on the COCO benchmark. The Yolov5's fast training and precision for small objects are due to daily improvements of the network, and the implementation of well-known architecture structures such as CSP-Darknet-53 \cite{CSPBackbone2020}, SPPF \cite{FSPP2015}, PANet feature extractor \cite{CSPPAN2018}, and a Yolov3 Head \cite{Yolov3}. We note that for the purpose of this work any off-the-shelf detection network with satisfying performances can be used instead of the YOLO network.

As the YOLOv5 network is pre-trained for object detection on general images, we employed transfer learning methods on training stage to get the adapted weights faster, with less data , and to leverage the existing object detection power of the YOLO network. The training has been done on a GPU. The GPU hardware is stated in section 5. Before training, the images were re-scaled into $640\times 480$ pixels, as every smartphone nowadays is capable of recording videos with this resolution, which is enough for the neural network to detect the tips of the drumsticks. The videos were randomly divided into two groups $85\%-15\%$, where the first group was used for training and the smaller group for validating. 

\section{Methodology} \label{sec:method}
\begin{figure}[h!]
\centering
\includegraphics[width=0.48\textwidth]{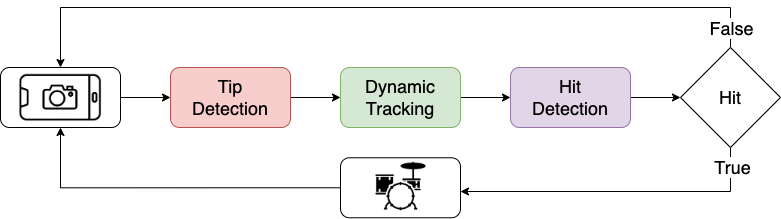}
\caption{The real-time detection and tracking algorithm.} \label{fig:algo}
\end{figure}

\begin{figure}
\centering
\includegraphics[width=0.49\textwidth]{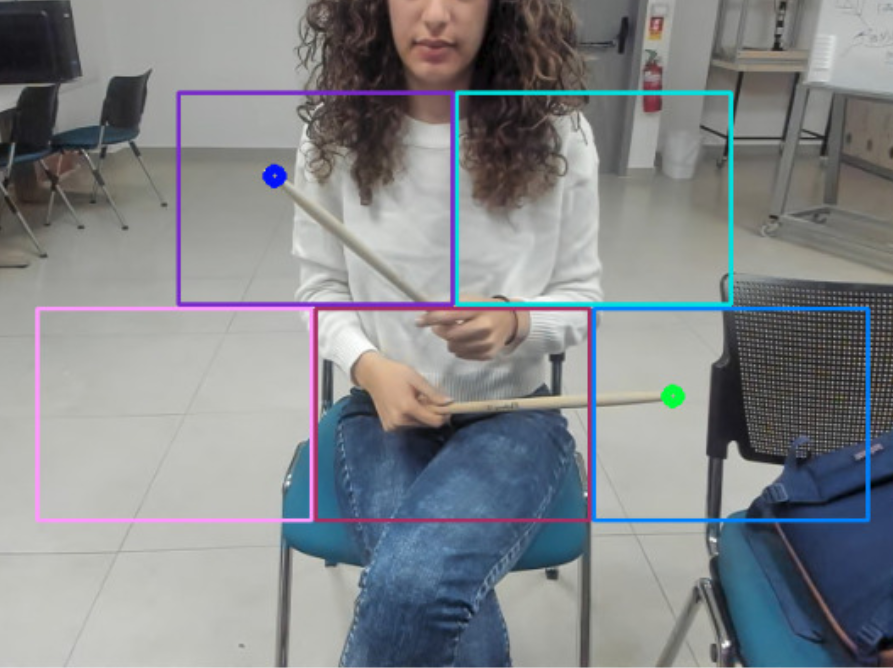}
\caption{Real-time detection of drum sticks, in a dynamic noisy environment. The five zones are marked with different colors to denote the type of drum in the set, and the tips are also colored differently to distinguish them.}\label{fig:drumdet}
\end{figure}

The method intends to imitate playing a drum set with multiple drums arranged before the player. Thus, the proposed method consists of two stages. First, the system initializes the acquisition frame. I.e., as seen in figure \ref{fig:drumdet}, a drum box boundary is set for each of the drums in the set. This indicates a zone in which, when the tip of the drumstick hits inside, the appropriate drum sound will be played. Note that there is no restriction on the zones being disjointed, thus, if a hit is made in a joint region between several zones, a composite sound will be played. Once the initialization is done, the real-time detection algorithm is invoked. The real-time algorithm is fed by frames acquired by the camera and reshaped to $640\times 480$ pixels resolution. This modest resolution allows for fast acquisition while maintaining a good enough resolution for detection. The algorithm itself is comprised of three main components. First, a detection algorithm is invoked to obtain an updated sample of tip positions. Then a tracking algorithm updates a dynamic belief on the position of the tips, and finally, a zero-crossing algorithm detects a hit. In case a hit is detected, the appropriate sound based on the hitting zone is played and the process repeats itself.
At any given time for user experience, the operating device renders on its display, the current frame with the zones marked and colored, differently for every type of drum, and the tips of the drumsticks as can be seen in figure \ref{fig:drumdet}.
In an air-playing setting, this is the only feedback and information about where every object is located the user receives.
An illustration of the algorithm is presented in figure \ref{fig:algo} Each of the colored blocks is discussed in detail in the following sections.

\subsection{Tips Detection}
\begin{figure}[h!]
\centering
\includegraphics[width=0.49\textwidth]{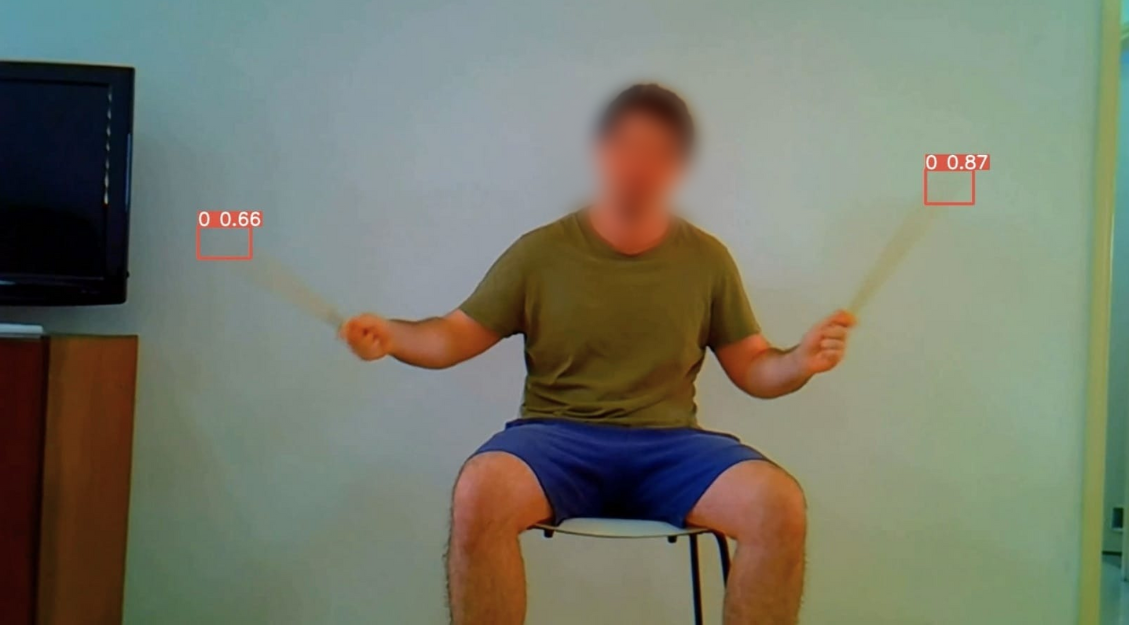}
\caption{Real-time tip detection under challenging conditions. The face was blurred for privacy purposes.} \label{fig:detect}
\end{figure}

In our system,YOLOV5 network receives a frame at 60 FPS, and at each acquisition time point it detects the tips in the frame. The (up to) three detections with the highest confidence are marked with the object's centroid and bounding box. The results are sent to a selection algorithm, which uses the network's prediction to determine which two objects are most probable according to the previous frame detection. The decision rule allows for stable detections if the previous frame contained close proximity detections or if the current frame contained far false-positive detections. 
The numeric information on locations and assignments is then passed on to the tracking algorithm. If the tips are not detected, the frame is discarded and the tracking algorithm estimates the positions of the tips as will be explained in section \ref{sec:kalman}. An example for the network detection abilities is presented in figure \ref{fig:detect}.

\subsection{Tip Tracking and Estimation} \label{sec:kalman}
While YOLO's accuracy is impressive compared to classical computer vision methods, it still suffers from two caveats. The first is that as with any object detection algorithm, YOLO is not foolproof. There are cases where the tips of the drumsticks are not detected, or only one of them is detected. The second is that YOLO only determines the tips' location within a given frame. Therefore, everything that happens between frames will elude the detection algorithm completely. I.e., a drum was hit between frames, resulting in two consecutive frames where the tip appears to be in the same place, while in fact, it dropped down to the drum, hit it, and came back up. As the frame rate of the camera is constrained, a method to estimate higher frequency events is required.

To this end we employ the celebrated Kalman filter \cite{Kalman1960}, and treat the problem as a tracking problem with disappearing targets \cite{weng2006video}. Thus, even when a position measurement has not been acquired, the filter may estimate the position of the desired target. Here, we make a relaxed assumption. While the drumsticks' movement is in three dimensions, most of the movement is in the plane in front of the player. Hence, we assume that displacement along the third axis can be neglected, and that the plane of movement is aligned with the camera frame. This assumption requires the player to place the camera right in front of him while using the system.

For simplicity we assume that the movement of the tips can be described by a linear second order model in each axis with a driving Gaussian noise as force input, and that the axes dynamics are not coupled. Also, in order to apply a tracking filter the system should be discretized. I.e., for a single drumstick and a single axis the governing dynamics is
\begin{equation}
    \begin{aligned}
        \frac{d^2}{dt^2} p(t) &= w(t) \\
        y(t) &= p(t) + v(t)
    \end{aligned} 
    \quad \Rightarrow \quad
    \begin{aligned}
        &x_{k+1} = Ax_k+Gw_k \\
        &y_k = Cx_k + v_k    
    \end{aligned}
\end{equation}
where $p(t)$ is the position, $w(t) \sim \mathcal{N}(0,\sigma_w)$ is the driving noise, and $v(t) \sim \mathcal{N}(0,\sigma_v)$ is the measurement noise, both assumed to be stationary white noises. 
In the discrete time system, $x_k=[p^x_k, v^x_k]^T$ is the state vector comprised of the position and velocity of the tip is the appropriate axis. $y_k$ is the position measurement at time $k$, $w_k$ is the driving force input, and $v_k$ is the measurement noise. We make further assumption and assume that the noise $w(t)$ is constant between every two consecutive time steps, i.e., $w(t) \equiv w(t_k)= w_k, \ t \in [t_k, t_{k+1}]$, hence the system matrices are
\begin{equation}
    \begin{aligned}
    &A = 
        \begin{bmatrix}
               1 & T \\
               0 & 1  \\
        \end{bmatrix}; \quad
    G = 
        \begin{bmatrix}
               \frac{T^2}{2} \\
               T \\
        \end{bmatrix}; \\
    &C = 
        \begin{bmatrix}
            1 & 0 & 0 & 0 
        \end{bmatrix};
    Q = 
    \begin{bmatrix}
        \frac{T^4}{4} & \frac{T^3}{2}  \\
        \frac{T^4}{2} & T^2 \\
    \end{bmatrix}\sigma_w
    \end{aligned}
\end{equation}
Here, $T$ is the sampling time step, which is a multiple of the frame rate. I.e., if it is equal to the frame rate the Kalman filter only compensates for the YOLO faults, if it is twice then also an intermediate time point is estimated between frames. We require T to be such that frames will necessarily be aligned by the filter time points.

\subsection{Hit detection algorithm}
Based on the mean square error estimate of the tips' positions, the hit detection algorithm determines if a hit occurred within a drum zone.
In that case, the detection algorithm maintains two FIFO queues of length N, for the previously estimated positions of the tips. In this work, N has been chosen to be 4. In order to correctly maintain the last N step trajectory of the tips, each queue is associated with a specific stick. Upon receiving an updated position estimate, the normal distance of the tip from each of the sticks is calculated and the estimate is associated with the closest tip. Then, a zero-crossing algorithm is employed for each of the queues, in order to detect a change of direction in the tip's movement. Since estimates might be noisy, a detection threshold was introduced to filter-out undesired zero-crossing, such as drumstick standing still, or unintentional jittery movement. If a hit is detected the appropriate sound is played according to the last frame zone, and the magnitude of the sound is linear with the velocity of the tip.

\section{Empirical Results} \label{sec:results}
The evaluation of our application has two aspects. The first aspect is a technical one, which purpose is to evaluate the qualitative performance of the system concerning the design requirements and objectives. I.e, the network detection performance, the tracking and estimation capabilities, etc. The second aspect is concerning the user experience of the system, e.g., interaction with the system, noticeable delays, etc.
All of the experiments of the research stage were conducted on the lab hardware, comprised of a standard Intel Core i7 PC with 8 cores, 64GB RAM and an NVIDIA GeForce RTX 3090 GPU.
In regards to camera quality, our study encompassed the utilization of two different camera systems based on the device being used: a dual 12MP camera system from a mobile device, and a webcam. Both cameras can handle 60 FPS. The training involved a balanced distribution of samples from both camera sources, maintaining a 50/50 ratio. Due to the conversion of training data to a mid resolution of 640x480 and the subsequent adjustment of input data to this resolution prior to inference, a high-resolution camera is not necessary, and thus a camera with moderate resolution capabilities will suffice for accurate results

%

\begin{figure*} [t]
\centering
\subfigure[Left hand.]
{
    \includegraphics[width=0.45\textwidth]{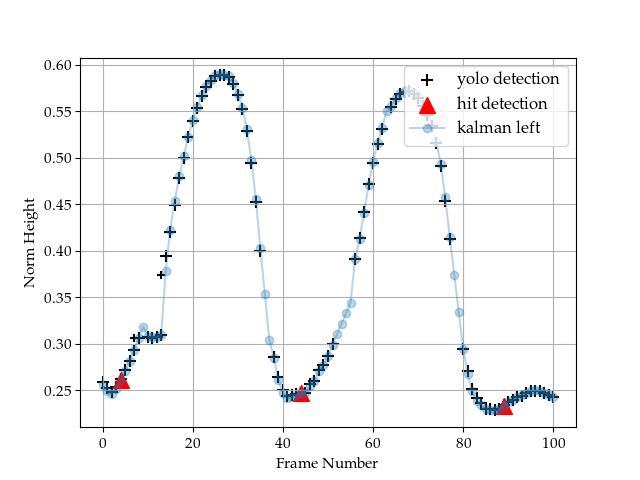}
    \label{fig:left}
}%
\subfigure[Right hand.]
{
    \includegraphics[width=0.45\textwidth]{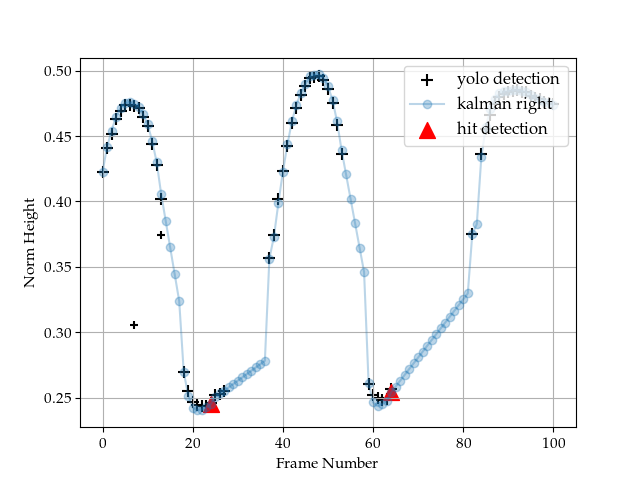}
    \label{fig:right}
}
\caption{Drumstick movements over 100 frames along the y-axis (up and down). The frames were taken from a detection of a drumstick striking 3 times with each hand, repeatedly, in 80 BPM.} \label{fig:left_right}
\end{figure*}

\subsection{Performance Results}
The first assessment was of the real-time detection performance of the overall system, compared to detection only by the detection network. This evaluation purpose is to test the robustness of the system to occlusions and miss detection and observe the estimation capabilities of the system. Figure \ref{fig:left_right} presents the results for this evaluation. Dark crosses mark the tip as detected by the detection network; for the frames, the detection was successful. The light circles are for the Kalman filter estimation of the location of the tip, and the red triangle marks a hit detection. In this case, the Kalman filter time step and the camera acquisition rate were matched. However, in general, the Kalman filter time step can be a multiple of the acquisition rate. As can be seen, the Kalman filter is able to filter out faulty detections, as can be seen in the first $20$ frames of figure \ref{fig:right}, and to successfully deal with tip occlusions, as can be seen from the missing detections in both sub-figures.

\begin{figure} [h!]
\centering
\includegraphics[width=0.45\textwidth]{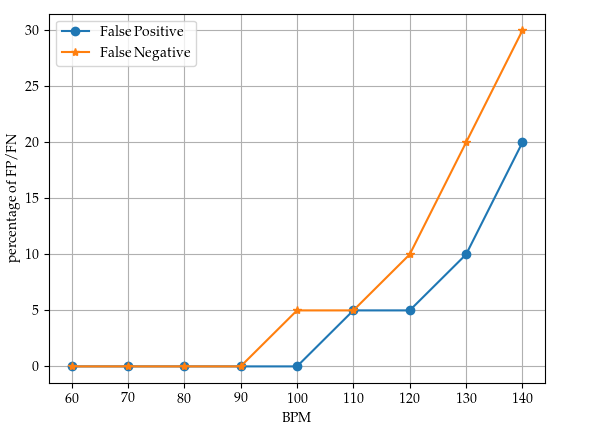}
\caption{FP and FN detections in tempos between 60 and 140 BPM.} \label{fig:FPFN}
\end{figure}

In addition, we defined two measures. The first, \textit{false positive} hits (FP), counts the number of hits that happened but the system failed to detect them. The second, \textit{false negative} hits (FN), counts the number of hits that the system wrongly detected (no actual hit). These two measures were tested against a range of beats per minute (BPM). For each test, the percentage of the measure is calculated after five minutes of continuous beating by the user. The results can be seen in figure \ref{fig:FPFN}.
\subsection{User Experience}
We evaluate our system qualitatively by conducting a survey taken from $10$ respondents. The respondents are categorized into three groups; beginners who have never played the drums, intermediates who played a few times, and advanced who play the drums for recreational purposes.
Each respondent played for five minutes and then asked to answer the following questions:
\begin{enumerate}
    \item How many times have you noticed a delay?
    \item How many times have the system missed a hit?
    \item How many times have the system sounded falsely?
    \item How will you rate your overall experience?
\end{enumerate}
In our survey $40\%$ of the respondents were beginners, $40\%$ were intermediates and $20\%$ were advanced players. The answers for the survey are reported in table \ref{tab:survey}.
 
\vspace{-1mm}

\renewcommand{\arraystretch}{1.5}
\begin{center}
\begin{table}[h!]
\centering
\begin{tabular}{c | c c c} 
 \ChangeRT{1mm}
 \textbf{Question\textbackslash Level} & \textbf{Beginner} & \textbf{Intermediate} & \textbf{Advanced} \\ [0.5ex] 
 \ChangeRT{1mm}
 Q1 & 10\% & 20\% & 35\% \\ 
 \hline
 Q2 & 10\% & 15\% & 25\% \\ 
 \hline
 Q3 & 10\% & 20\% & 25\% \\ 
 \hline
 Q4 & 95\% & 90\% & 80\% \\ 
 \ChangeRT{1mm}
\end{tabular}
\vspace{3mm}
\caption{Results of the user experience questionnaire.}
    \label{tab:survey}
\end{table}
\end{center}
\vspace{-2mm}
The answers show that as user expertise rises, it also becomes more sensitive to delays and misses. Overall system satisfaction declines as a consequence. The results indicate that the goal of designing a system suitable for recreational purposes with high satisfaction, specifically for beginners to intermediate users has been met. Moreover, even advanced users find the system acceptable with an overall rating of $80\%$.

 
\section{Conclusion and Future Work} \label{sec:conc}
In this work, we have presented a proof-of-concept (POC) for an affordable anywhere anytime system for playing drums, requiring only a standard camera and computation capabilities, found nowadays in almost every mobile device. We have presented quantitative results for the performance of each of the components in the system and evaluated its robustness and confidence. In addition, we have presented initial qualitative results from users about the user experience of the entire system that serve as motivation and verification of success to merit future research. In future work, there are few directions to follow. One can focus on investigating the tactile feedback and explore the difference between playing on air or on different materials. In another direction, we intend to collect a more diverse dataset to represent reliably the user candidates of the system. Furthermore, we intend to develop and optimize the current POC into a fully functional prototype that can be tested under real hardware conditions.

\bibliographystyle{IEEEbib}
\bibliography{Bibliography}

\end{document}